\documentclass[runningheads,citeauthoryear]{apinv}
\usepackage{epsfig,cite,graphics}
\usepackage{marvosym} 
\usepackage[utf8]{inputenc}
\usepackage{tabu}
\usepackage{url}

\begin{document}

\title{Lightcurve and rotation period \\
determination for asteroid 3634 Iwan}
\titlerunning{Lightcurve and rotation period for 3634 Iwan}
\author{Z. Donchev\inst{1}, E. Vchkova Bebekovska\inst{2}, A. Kostov\inst{1} and G. Apostolovska\inst{2}}
\authorrunning{Z. Donchev et al.}
\tocauthor{Z. Donchev} 
\institute{
\inst{1}Institute of Astronomy and National Astronomical Observatory,  Bulgarian Academy of Sciences, Tsarigradsko Chaussee Blvd. 72, BG-1784, Sofia, Bulgaria
\newline
\inst{2}Institute of Physics, Faculty of Science, Ss. Cyril and Methodius University, Skopje, Republic of Macedonia, Arhimedova 3, 1000 Skopje, Republic of Macedonia	
\newline
	\email{akostov@astro.bas.bg}    }
\papertype{Submitted on 15.05.2018; Accepted on 04.06.2018}	
\maketitle

\begin{abstract}
Lightcurve of the asteroid 3634 Iwan observed at the Bulgarian National Astronomical Observatory Rozhen in 2017 is presented. The asteroid was observed only one night, on 22 March 2017 accidentally in the field of view in which our target for shape modeling asteroid 289 Nenetta was positioned. The Asteroid Lightcurve Database (LCDB; Warner et al. 2009) did not contain any previously reported results for 3634 Iwan. If we accept that the ligtcurve of the asteroid is typical with two maxima and minima, a single night observations covers the whole rotational cycle. The lightcurve plotted by MPO Canopus provides a best fit to synodic period of $4.72\pm 0.06$ h with amplitude of $0.15\pm 0.02$ mag. Our assessment is that for the uniqueness of the period solution we need longer observational span which will reveal the number of extrema of the lightcurve and confirm the quality code U=3.
\end{abstract}
\keywords{Minor planets, asteroids, photometric-Asteroids: individual: 3634 Iwan}

\section{Introduction}

Knowledge of the dynamical  and physical characteristics of asteroids brings researchers close to understanding past and present processes of our planetary system. Asteroids or minor planets, since they have no changes or very small changes of the original material from which they were formed,  are carrying crucial information about the Solar System formation and its evolution.  Beside the scientific aspect, the asteroids research is important for humanity for many reasons. One of this is discovering of potentially hazardous asteroids and prevent an unwanted accident. The most practical reason is  the utilization of asteroids as sources of rare minerals and metals. Last year, asteroid exploration gained a new scientific significance and excitement with the discovery of the first asteroid that has entered the Solar System from interstellar space (Meech et al. 2017).  This interstellar object, previously known as A/2017 U1, was named 'Oumuamua – a Hawaiian-based word meaning a messenger reaching out from the distant past. Because astronomers believe that many such asteroids enter in our Solar System from interstellar space, IAU introduced a new cataloguing system for interstellar asteroids and 'Oumuamua was designated as I1/2017 U1, with the "I" for interstellar and "1" because it is the first.

Due to the huge number of asteroids and limited opportunities to visit them by spacecraft for a detailed study, information about morphology and dynamic characteristics of asteroids are obtained from ground-based telescopes. The most common observations of the asteroids are the observations based on photometry. In order to get more precise results these observations are often combined with spectroscopic, polarimetric, thermal IR and radar observations. The photometric measurements of the variation of the total brightness of the asteroids as a function of time give the lightcurve which reveals the rotation period of the asteroid and can carries information about the spin poles and shape of the asteroid. For very limited number of asteroids using interferometry, adaptive optics or occultation, obtained disk-resolved data could give more precise shape details. Asteroids which are our targets for shape modelling are chosen to have rotation period between  three and six hours, in order the whole synodic period to be covered, during one observational night. To solve the inverse problem of reconstructing the 3D  shape model and spin axis direction from a set of lightcurves we have to observe the asteroid at diferrent geometric conditions during several oppositions (Kaasalainen and Torppa (2001) and Kaasalainen et al. (2001)).

\section{Observations and data reduction}

The photometric observation of 3634 Iwan was performed at the BNAO Rozhen by 50/70 Schmidt telescope with an FLI PL 16803 CCD  with $4096\times4096$ array of $9\mu m$ square pixels. All images were taken through an R filter. We performed standard image reduction using dark frame subtraction and flat-field correction. For aperture photometry of the asteroid and the comparison stars  we used CCDPHOT by Buie (1996). Lightcurve analysis (composite lightcurve, synodic rotational period and estimation of the amplitude of the ligtcurve) were performed using MPO Canopus v10.7.7.0\footnote{MPO Canopus Software: \url{http://www.MinorPlanetObserver.com}} (Warner 2016).

In Table \ref{table1} the aspect data for 3634 Iwan for the night of observations are reported. In the first column is the name of the asteroid followed by the date of the observation referring to the mid-time of the lightcurve observed, asteroid distance from the Sun ($r$), from the Earth ($\Delta$), the solar phase angle, and the J2000.0 ecliptic longitude ($\lambda$) and latitude ($\beta$) of the asteroid referred to the time in the second column.

\begin{table}[h]
\begin{center}
\caption{Aspect data}
\setlength{\tabcolsep}{6pt}
\setlength{\extrarowheight}{1.5pt}
\begin{tabu}{ccccccc}
\hline
\\[-0.295cm]
Asteroid & Date & $r$ & $\Delta$ & Phase angle & $\lambda$ & $\beta$ \\
         & (UT) & (AU) & (AU) & ($^\circ$) & ($^\circ$) & ($^\circ$) \\[0.095cm]
\hline
\\[-0.295cm]
3634 Iwan & 2017 March 22.96 & 2.3602	& 1.3806 & 5.84	& 168.64 & -1.60 \\[0.030cm]
\hline
  \end{tabu}
  \label{table1}
  \end{center}
\end{table}

\section{Results}

Asteroid 3634 Iwan is a main belt asteroid and it orbits around the Sun for the  period of 3.36  years. It was discovered by Swedish astronomer Claes-Ingvar Lagerkvist on March 16, 1980 at the La Silla observatory (Schmadel 1997). The discoverer named this asteroid in honor of Iwan P. Williams of Queen Mary College, London, with whom he had long and fruitful collaboration. Iwan P. Williams is in the list of astronomers who are credited by the Minor Planet Center\footnote{\url{https://www.minorplanetcenter.net/iau/lists/MPDiscsNum.html}} with discovery of one or several minor planets and, in this case, in recognition of his well-known work on meteor streams and investigations of comets.

\begin{figure}[h]
\begin{center}
	\centering{\epsfig{file=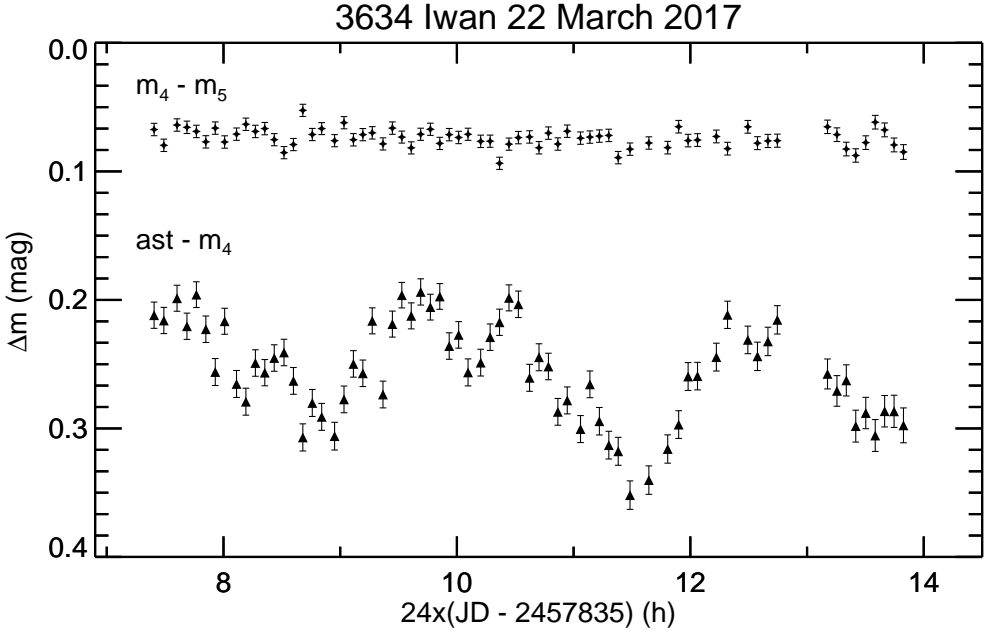, width=0.99\textwidth}}
	\caption{The relative lightcurve of 3634 Iwan on 22 March 2017.}
	\label{fig1}
\end{center}
\end{figure}

On 22 March of 2017 on Schmidt telescope we observed the asteroid 289 Nenetta which was our target for shape modeling. The field of view was observed about six and half hours which almost covered one rotational period of our target. We used exposure time of 270s for 289 Nenetta which was about 14.5 mag.  In the same images beside 289 Nenetta, we noticed  during reduction of the data, a very faint asteroid with 16.7 mag, which was identified as 3634 Iwan. Till now this asteroid has no published lightcurve or rotation period. The opposition of 3634 Iwan in 2017 was on March 11 and we observed the asteroid soon after its opposition at the solar phase angle of $5.8^{\circ}$. The individual relative ligtcurve is presented in Fig.\ref{fig1}.

\begin{figure}[h]
\begin{center}
	\centering{\epsfig{file=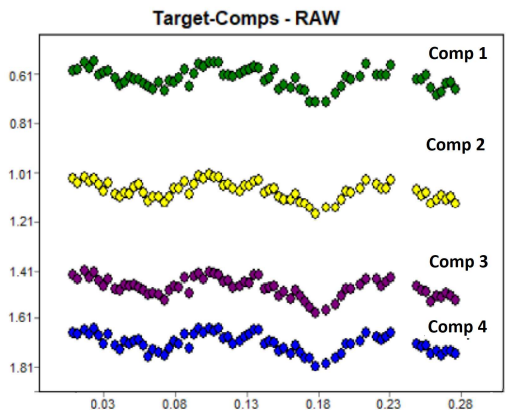, width=0.5\textwidth}}
	\caption{Derived lightcurves (plot of magnitude vs. time in days) of 3634 Iwan when using 4 different comparison stars on 22 March 2017.}
	\label{fig2}
\end{center}
\end{figure}

\newpage

\begin{figure}[h]
\begin{center}
	\centering{\epsfig{file=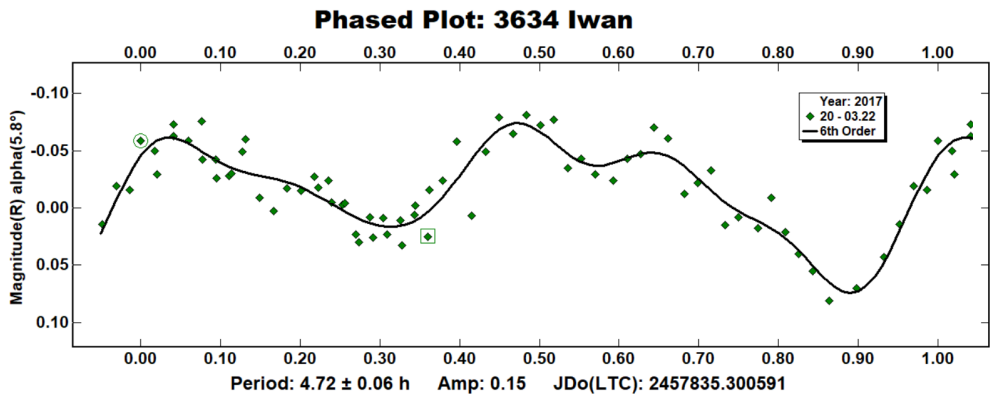, width=0.99\textwidth}}
	\caption{Lightcurve that provides a best fit to $4.72\pm 0.06$  h with amplitude of $0.15\pm 0.02$ mag.}
	\label{fig3}
\end{center}
\end{figure}

In Fig.\ref{fig3} we present a lightcurve plotted by MPO Canopus that provides a best fit to $4.72\pm 0.06$  h as is using 6-order fit Fourier analysis. The RMS scatter of the fit is 0.017 mag and amplitude of the Fourier model curve is $0.15\pm 0.02$ mag. The x-axis rotational phase ranges from –0.05 to 1.05, and the magnitudes were normalized to the phase angle given in parentheses using G = 0.15.

\vspace{-0.7cm}

\begin{figure}[h]
\begin{center}
	\centering{\epsfig{file=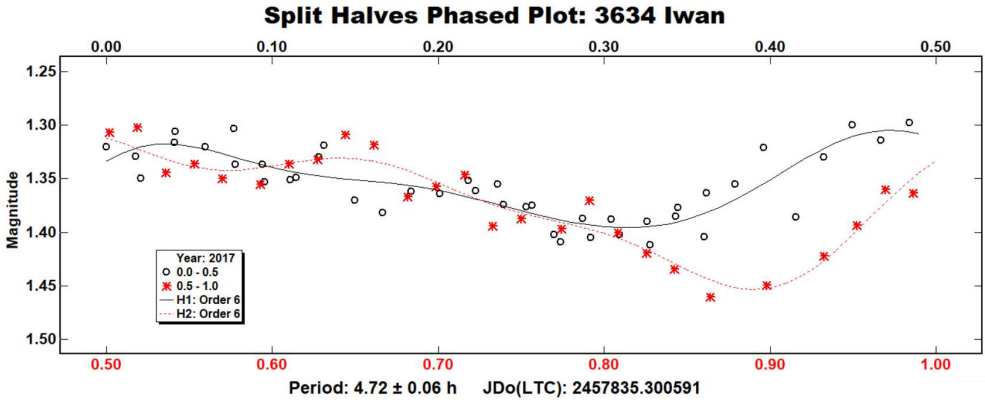, width=0.99\textwidth}}
	\caption{3634 Iwan split-halves phased plot for 4.72 h shows the inequality of two halves.}
	\label{fig4}
\end{center}
\end{figure}

We used split-halves utility in MPO Canopus, where the second half of the lightcurve is superimposed on top of the first half. Having in mind that obtained relative LC reveals two peaks with different heights and not equal widths two halves are far from identical (see Fig.\ref{fig4}).

\begin{figure}[h]
\begin{center}
	\centering{\epsfig{file=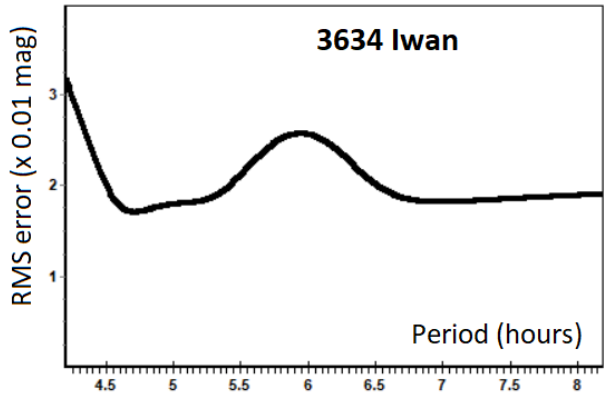, width=0.6\textwidth}}
	\vspace{0.5cm}
	\caption{The period spectrum for 3634 Iwan based on our observations from 2017.}
	\label{fig5}
\end{center}
\end{figure}

Period analysis was done using MPO Canopus, which incorporates the Fourier analysis algorithm (FALC) developed by Harris et al. (1989). The estimated rotational period is $4.72\pm 0.06$ h which has a smallest value of RMS error (Fig.\ref{fig5})  and best correspondence to the observational points.

\vspace{-0.7cm}

\begin{figure}[h]
\begin{center}
	\centering{\epsfig{file=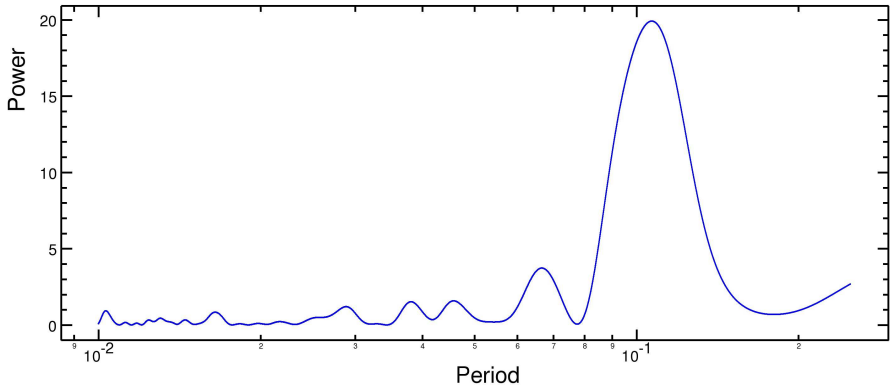, width=0.89\textwidth}}
	\caption{The NASA Exoplanet Archive system periodogram for our data.The highest peak reveals the period P=0.10679196 days (2.563h) which  coresponds to one half of the assumed period.}
	\label{fig6}
\end{center}
\end{figure}

We tried to calculate the period using the Periodogram service in NASA Exoplanet Archive system\footnote{\url{https://exoplanetarchive.ipac.caltech.edu/index.html}}. We chose Lomb-Scargle (Scargle 1982) algorithm which is an approximation of the Fourier Transform for unevenly spaced time sampling. The maximum periodogram power value  is presented in Fig.\ref{fig6}.

\newpage

\begin{figure}[h]
\begin{center}
	\centering{\epsfig{file=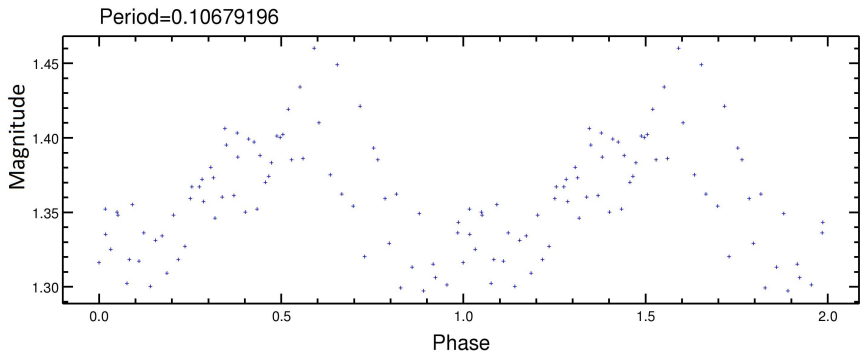, width=0.86\textwidth}}
	\caption{The LC with Period = 0.10679196 days obtained from the periodogram from NASA Exoplanet Archive system.}
	\label{fig7}
\end{center}
\end{figure}

The obtained LC in Fig.\ref{fig7} is similar to the split-halves utility in MPO Canopus in the case where the period is taken double of 0.10679196 days (or equal to 5.126h) and the phase 1 actually corresponds to half of the phase. Knowing that obtained relative LC reveals  two peaks with different heights and not equal widths we can not take double of this period as the precise synodic period of rotation of the asteroid. Thus we can conclude that the first and deepest minimum in period spectrum in Fig.\ref{fig5}, reveals the most appropriate value of the rotation period of 3634 Iwan. Additional data obtained in longer time span of observation could show more complex shape of the lightcurve and the more accurate value of the period.


\newpage

\section*{Conclusion}
The rotational period of 3634 Iwan of  $4.72\pm 0.06$ h and amplitude of $0.15\pm 0.02$ mag is calculated for the first time. Bearing in mind that the ligtcurve of this asteroid might be no typical with only two extrema we can not completely accept the uniqueness of obtained period solution. In order to calculate the period with greater precision we need observations from aditional nights. According to the visibility of 3634 Iwan and the availability of the telescope at NAO Rozhen the asteroid is planned for observations in its next apparitions.

\section*{Acknowledgements}
Authors gratefully acknowledge the observing grant support from the Institute of Astronomy and Rozhen National Astronomical Observatory, Bulgarian 
Academy of Sciences.

This research has made use of the NASA Exoplanet Archive, which is operated by the California Institute of Technology, under contract with the National Aeronautics and Space Administration under the Exoplanet Exploration Program.

\end{document}